\documentclass{elsart41}
\usepackage{graphics}
\usepackage{graphicx}
\usepackage{amssymb}
\begin{document}

\begin{frontmatter}



\title{Novel  quantum criticality  
due to  emergent  topological conservation law in  high-$T_c$ cuprates
}
%

\author{T.K. Kope\'{c}\corauthref{Name1}}

\address{Institute for Low Temperature and Structure Research,
Polish Academy of Sciences, P.O.Box 1410, 50-950 Wroclaw 2, Poland}  

\corauth[Name1]{e-mail: kopec@int.pan.wroc.pl}

\begin{abstract}
We argue that in strongly correlated
electron  system  collective  instanton excitations of the phase field (dual to the charge)
arise with a great degree of stability, governed by gauge flux changes by an integer multiple of $2\pi$. By unraveling consequences of the   non-trivial topology of the charge  gauge U(1) group, we found that the pinning of $\mu$ and the zero-temperature divergence of charge compressibility $\kappa\sim\partial n_e/\partial\mu$  defines
novel ``hidden" quantum criticality on verge of the Mott transition governed by 
the protectorate of stable topological numbers rather than Landau paradigm of  the symmetry breaking.
\end{abstract}

\begin{keyword}
strongly correlated systems \sep superconductivity \sep topological order
\PACS    74.20.-z, 74.20.Mn, 74.72.-h
\end{keyword}
\end{frontmatter}
The understanding of the observed doping dependence
of $\mu$  in  cuprates remains a puzzle.
In the case of LSCO, the angle-resolved photoemission
spectroscopy  studies \cite{chem1} have shown that in underdoped
samples, the chemical potential  is pinned 
above the top of the lower Hubbard band.
The photoemission
measurements of core levels also shows that $\mu$
does not move with hole doping in the underdoped
region. The  behavior of $\mu$
is quite peculiar from the viewpoint of the Fermi liquid theory of the metallic states and signals
a dramatic reorganization of the electronic structure of cuprates  with  doping.
In cuprates there is clear evidence for the existence of a special
doping point in the lightly-overdoped region  where superconductivity is
most robust. This indicates that it could be a  quantum critical point
(QCP) while  the  critical fluctuations might be responsible for the 
unconventional normal state behaviour \cite{qcp}.
The resemblance to a conventional QCP is hampered by the lack of any clear signature of
thermodynamic critical behavior.
Experiments appear to exclude any broken  symmetry 
around this point  although  a sharp change in transport properties is 
observed \cite{transport}.
We explore Mott transitions from  the  non-magnetic insulator to a superconductor induced by doping and show that the process is governed by the  topological structure of the electromagnetic compact gauge U(1) group. As a result  collective 
instanton excitations of the phase field (dual to the charge) arise with a great degree of stability, governed by gauge flux changes by an integer multiples of $2\pi$,
which  labels topologically ordered ground states. The associated abrupt transition
between differnt ``vacua" allows us to make link between the
unusual behavior of the chemical potential and a novel type of
quantum criticality that goes beyond the paradigm of the  symmetry breaking.

We consider an effective one--band electronic
Hamiltonian on a tetragonal lattice that emphasises strong anisotropy and
the presence of a layered CuO$_2$ stacking sequence in cuprates:
$H=\sum_\ell H^{(\ell)}$, where
$H^{(\ell)}= {H}_{tJ}^{(\ell)}+{ H}_{U}^{(\ell)}+{H}_\perp^{(\ell)}+H_\mu^{(\ell)}$, where
\begin{eqnarray}
&&{H}_{tJ}^{(\ell)}= -\sum_{\langle {\bf r}{\bf r}'\rangle}
 tc^{\dagger }_{{\alpha}\ell}({\bf r})
c_{\alpha \ell}({\bf r}')
+t'\sum_{{\langle \langle{\bf r}{\bf r}'\rangle\rangle}}
 c^{\dagger }_{{\alpha}\ell}({\bf r})
c_{\alpha \ell}({\bf r}') 
\nonumber\\
&&+J\sum_{{\langle {\bf r}{\bf r}'}\rangle}
\left[{\bf S}_\ell{({\bf r})}
\cdot{\bf S}_\ell{({\bf r}')}
-\frac{{n}_\ell{({\bf r})}{n}_\ell{({\bf r}')}}{4} \right].
\label{tj}
\end{eqnarray}
Here, $\langle {\bf r},{\bf r}'\rangle$ and 
$\langle\langle {\bf r},{\bf r}'\rangle\rangle$
denotes  summation
over the nearest-neighbour and next--nearest--neighbour
sites labelled by $1\le {\bf r}\le N$ within the CuO  plane, respectively, with $t,t'$ being the {\it bare}  hopping integrals  $t'>0$, while $1\le\ell\le N_\perp$ labels copper-oxide layers.
The operator $c_{\alpha\ell}^\dagger({\bf r})$
creates  an electron of spin $\alpha$ at the lattice site $({\bf r},\ell)$. Next,
$S^a_{\ell}({\bf r})$  stands for spin operator and  $J$ is the antifferomagnetic exchange.
Further, ${n}_\ell({{\bf r}})= n_{\uparrow\ell} ({\bf r})+n_{\downarrow\ell}({\bf r})$ is 
the electron number operator, where 
${n}_{\alpha\ell}({{\bf r}})= c^\dagger_{\alpha\ell}({{\bf r}})
c_{\alpha\ell}({\bf r})$, respectively; $H_\mu^{(\ell)}=-\mu\sum_{{\bf r}}{n}_\ell{({\bf r})}$	
 and $\mu$ is the chemical potential.
The Hubbard term is ${H}_U^{(\ell)}=\sum_{\ell\bf r}
Un_{\uparrow\ell} ({\bf r}) n_{\downarrow\ell}({\bf r})$ with the on--site repulsion Coulomb energy $U$,
while ${H}_\perp^{(\ell)}=-
\sum_{{\bf r}{\bf r}'}
t_\perp({\bf r}{\bf r}')
 c^{\dagger }_{{\alpha}\ell}({\bf r})
c_{\alpha \ell+1}({\bf r}')$ facilitates the interlayer coupling,
where $t_\perp$ is the interlayer hopping
with the $c-$axis dispersion  $\epsilon_\perp({\bf k},k_z)=2t_\perp({\bf k})\cos(ck_z)$,
while $t_\perp({\bf k})=t_\perp\left[\cos(ak_x)-\cos(ak_y)\right]^2$.

We decouple the Hubbard term $H_U$ using  the collective variable and
 $iV({\bf r}\tau)$ 
conjugate to the local particle number $n_\ell({\bf r}\tau)$.
Further, we introduce the {\it phase } (or ``flux") field ${\phi}_\ell({\bf r}\tau)$ via the Faraday--type relation 
$\dot{\phi}_\ell({\bf r}\tau)\equiv\frac{\partial\phi_\ell({\bf r}\tau)}
{\partial\tau}=\tilde{V}_\ell({\bf r}\tau)$
and perform the  gauge transformation to the {\it new} fermionic variables $f_{\alpha\ell}({\bf r}\tau)$, where
$c_{\alpha\ell}({\bf r}\tau)=e^{i{\phi}_\ell({\bf r}\tau)}f_{\alpha\ell}({\bf r}\tau)$.
The electromagnetic  U(1)
group governing the phase field is {\it compact}, {\it i.e.}
$\phi_\ell({\bf r}\tau)$  has the topology of a circle ($S_1$), so that
{\it instanton } effects can arise due to  non-homotopic 
mappings of the configuration space onto the gauge group $S_1\to$ U(1).
Therefore, we concentrate on 
closed paths  in the imaginary time $0\le\tau\le\beta\equiv1/k_BT$ 
 which fall into distinct, disconnected (homotopy) classes
labelled by the integer  winding number $m_\ell({\bf r})$\cite{schulman}. 
In the limit of strong (weak) correlations the electron  number $n_e\equiv\langle \bar{c}c\rangle$ interpolates between topological $n_b\equiv\langle m\rangle$ (fermionic $n_f\equiv\langle \bar{f}f\rangle$) occupation numbers \cite{kopec}.
In the large--$U$ limit 
$\mu\to n_fU/2$, so that $n_e\to n_b$  and the system behaves as governed entirely by
U(1) topological charges which play the role of ``quasiparticles".
Moreover, due to the  frustrated
motion of the carriers in the fluctuating bath of U(1) gauge potentials
the actual tight-binding parameters become ``dressed" 
$t^\star_X=t_X \langle e^{-i[\phi_\ell({\bf r}\tau)-\phi_\ell({\bf r}'\tau)]}\rangle$, where 
$t_X=t,t',t_\perp$ are the bare band parameters. 
It is instructive to calculate the charge compressibility
 $\kappa=\partial n_e/\partial\mu$. The result is given
in Fig.1 along with the outcome for the superconducting phase boundary.
We see the evolution of $\kappa$ with decreasing $n_e$, (i.e. hole doping) from the Mott insulator \cite{mott}
with $\kappa=0$ (at $2\mu/U=1$) to a point of degeneracy  on the brink of the
particle occupation change at $2\mu/U=.5$ where $\kappa=\infty$ at $T=0$.
This is also the point on the phase diagram   from which the superconducting lobe emanates.
It is clear that, the nature of the divergence of $\kappa$ here has little to do
with singular fluctuations due to spontaneous symmetry breaking as in the ``conventional" phase transition.
Rather, this divergent response appears as a kind of topological protection
built in the system against
the small changes of $\mu$.
Further, $\kappa\to \infty$ implies that the and $\partial\mu/\partial n_e$
becomes vanishingly small  at $T=0$ which results in the  chemical potential pinning 
(see, inset of Fig.1).

To conclude,
topological effects arise as stable, non-perturbative, collective excitations of the phase field
(dual to the charge), which  carry  novel topological characteristics.
These are the winding numbers of U(1) group:
$ m_\ell({\bf r})\equiv\frac{1}{2\pi}\int_0^\beta d\tau\dot{\phi}_\ell({\bf r}\tau)$ that
 become topologically conserved quantities.
It is exactly the appearance of these topological charges
that render the system protected against small changes of the Hamiltonian's
parameters.
This novel conservation does not arise  just out of a
symmetry of the theory (as ``conventional" conservation laws based on Noether's theorem) 
but it is a consequence of the connectedness, i.e. topology 
of the phase  space,  related to the topological properties of the associated symmetry group.
\begin{figure}[!t]
\begin{center}
\includegraphics[width=0.45\textwidth]{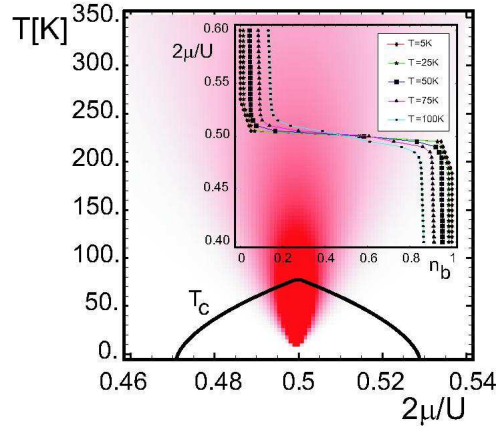}
\end{center}
\caption{The superconducting critical temperature $T_c$ as a function of
the chemical potential $\mu$ together with the density plot of the  charge  compressibility
$\tilde {\kappa}\equiv U\kappa/2$ for $t^\star=0.5$eV, $t'^\star/t^\star=0.3$, $t^\star_\perp=0.01$eV,
 $J=0.15$eV and $U=4$eV. Inset: the chemical potential $\mu$ as a function of the
occupation number $n_e\approx n_b$ for various temperatures $T$. The value of $\mu$ stays within the charge gap
as  $n_b$ changes. 
}
\label{fig1}
\end{figure}


\begin{thebibliography}{99}
\bibitem{chem1}
%
 A. Ino et al, Phys.
Rev. B {\bf 62}  (2000) 4137.
%
\bibitem{qcp}
C.M.Varma, Phys. Rev. B{\bf  55} (1997) 14554.
%
\bibitem{transport}
T. Ito, K. Takenaka, and S. Uchida, Phys. Rev. Lett. {\bf 70} (1993) 3995.
%
\bibitem{schulman}
L. S. Schulman, {\it Techniques and Applications of Path
Integration} (Wiley, New York 1981). 
%
\bibitem{kopec}
T. K. Kope\'c, Phys. Rev. B{\bf 70} (2004) 054518.
%
\bibitem{mott}
N. F. Mott, {\it Metal-Insulator Transitions} (Taylor \& Francis, London, 1990).
%
\end{thebibliography}
\end{document}